\documentclass[english,aps,prl,reprint,superscriptaddress,numerical,showpacs,floatfix,nofootinbib,twocolumn]{revtex4}
\usepackage[T1]{fontenc}
\usepackage[latin9]{inputenc}
\usepackage[pdftex]{color}
\usepackage{babel}
\usepackage{amssymb}
\usepackage[pdftex]{graphicx}
\usepackage{tabularx}
\usepackage{footnote}
\usepackage{mathrsfs}
\usepackage{physics}
\usepackage[unicode=true,pdfusetitle,
 bookmarks=true,bookmarksnumbered=false,bookmarksopen=false,
 breaklinks=false,pdfborder={0 0 1},backref=false,colorlinks=true]
 {hyperref}
\hypersetup{
 urlcolor=blue, citecolor=blue}

\usepackage{dcolumn}
\newcolumntype{d}{D{.}{.}{1} }

\makeatletter

\@ifundefined{textcolor}{}
{%
 \definecolor{BLACK}{gray}{0}
 \definecolor{WHITE}{gray}{1}
 \definecolor{RED}{rgb}{1,0,0}
 \definecolor{GREEN}{rgb}{0,1,0}
 \definecolor{BLUE}{rgb}{0,0,1}
 \definecolor{CYAN}{cmyk}{1,0,0,0}
 \definecolor{MAGENTA}{cmyk}{0,1,0,0}
 \definecolor{YELLOW}{cmyk}{0,0,1,0}
}

\hypersetup{urlcolor=blue}

\makeatother

\begin{document}

\title{Precision mass measurements of magnesium isotopes and implications on the validity of the Isobaric Mass Multiplet Equation}

\author{M.~Brodeur}
\email{mbrodeur@nd.edu}
\affiliation{Department of Physics, University of Notre Dame, Notre Dame, IN 46556 USA}
\author{A.A.~Kwiatkowski}
\affiliation{TRIUMF, Vancouver, BC V6T 2A3 Canada}
\affiliation{Department of Physics, Texas A$\&$M University, College Station, TX 77843 USA}
\author{O.~M.~Drozdowski}
\affiliation{TRIUMF, Vancouver, BC V6T 2A3 Canada}
\affiliation{Institute for Theoretical Physics, Universit\"at Heidelberg, Philosophenweg 12, D-69120 Germany}
\author{C.~Andreoiu}
\affiliation{Department of Chemistry, Simon Fraser University, Burnaby, BC V5A 1S6 Canada}
\author{D.~Burdette}
\affiliation{Department of Physics, University of Notre Dame, Notre Dame, IN 46556 USA}
\author{A.~Chaudhuri}
\altaffiliation[Present address: ]{Canadian Nuclear Laboratories, Chalk River, ON K0J 1J0, Canada}
\affiliation{TRIUMF, Vancouver, BC V6T 2A3 Canada}
\author{U.~Chowdhury}
\affiliation{TRIUMF, Vancouver, BC V6T 2A3 Canada}
\affiliation{SNOLAB, Lively, ON P3Y 1N2 Canada}
\affiliation{Department of Physics and Astronomy, University of Manitoba, Winnipeg, MB R3T 2N2 Canada}
\author{A.T.~Gallant}
\affiliation{TRIUMF, Vancouver, BC V6T 2A3 Canada}
\affiliation{Department of Physics and Astronomy, University of British Columbia, Vancouver, BC V6T 1Z1 Canada}
\affiliation{Physical and Life Science Directorate, Lawrence Livermore National Laboratory, Livermore, CA 94550 USA}
\author{A.~Grossheim}
\affiliation{TRIUMF, Vancouver, BC V6T 2A3 Canada}
\author{G.~Gwinner}
\affiliation{Department of Physics and Astronomy, University of Manitoba, Winnipeg, MB R3T 2N2 Canada}
\author{H.~Heggen}
\affiliation{TRIUMF, Vancouver, BC V6T 2A3 Canada}
\author{J.~D.~Holt}
\affiliation{TRIUMF, Vancouver, BC V6T 2A3 Canada}
\author{R.~Klawitter}
\affiliation{TRIUMF, Vancouver, BC V6T 2A3 Canada}
\affiliation{Max-Planck-Institut f\"{u}r Kernphysik, D-69117 Heidelberg, Germany}
\author{J.~Lassen}
\affiliation{TRIUMF, Vancouver, BC V6T 2A3 Canada}
\affiliation{Department of Physics and Astronomy, University of Manitoba, Winnipeg, MB R3T 2N2 Canada}
\author{K.G.~Leach}
\affiliation{TRIUMF, Vancouver, BC V6T 2A3 Canada}
\affiliation{Department of Chemistry, Simon Fraser University, Burnaby, BC V5A 1S6 Canada}
\affiliation{Department of Physics, Colorado School of Mines, Golden, CO 80401 USA}
\author{A.~Lennarz}
\affiliation{TRIUMF, Vancouver, BC V6T 2A3 Canada}
\affiliation{Institut f\"{u}r Kernphysik, Westf\"{a}lische Wilhelms-Universit\"{a}t, D-48149 M\"{u}nster, Germany}
\author{C.~Nicoloff}
\affiliation{Department of Physics, University of Notre Dame, Notre Dame, IN 46556 USA}
\affiliation{Department of Physics, Wellesley College, Wellesley, MA 02481 USA}
\author{S.~Raeder}
\affiliation{TRIUMF, Vancouver, BC V6T 2A3 Canada}
\author{B.E.~Schultz}
\affiliation{TRIUMF, Vancouver, BC V6T 2A3 Canada}
\author{S.~R.~Stroberg}
\affiliation{TRIUMF, Vancouver, BC V6T 2A3 Canada}
\author{A.~Teigelh\"{o}fer}
\affiliation{TRIUMF, Vancouver, BC V6T 2A3 Canada}
\affiliation{Department of Physics and Astronomy, University of Manitoba, Winnipeg, MB R3T 2N2 Canada}
\author{R.~Thompson}
\affiliation{Department of Physics and Astronomy, University of Calgary, AB T2N 1N4 Canada}
\author{M.~Wieser}
\affiliation{Department of Physics and Astronomy, University of Calgary, AB T2N 1N4 Canada}
\author{J.~Dilling}
\affiliation{TRIUMF, Vancouver, BC V6T 2A3 Canada}
\affiliation{Department of Physics and Astronomy, University of British Columbia, Vancouver, BC V6T 1Z1 Canada}

\date{\today}
\begin{abstract}
If the mass excess of neutron-deficient nuclei and their neutron-rich mirror partners are both known, it can be shown that  deviations of the Isobaric Mass Multiplet Equation (IMME) in the form of a cubic term can be probed. Such a cubic term was probed by using the atomic mass of neutron-rich magnesium isotopes measured using the TITAN Penning trap and the recently measured proton-separation energies of $^{29}$Cl and $^{30}$Ar. The atomic mass of $^{27}$Mg was found to be within 1.6$\sigma$ of the value stated in the Atomic Mass Evaluation. The atomic masses of $^{28,29}$Mg were measured to be both within 1$\sigma$, while being 8 and 34 times more precise, respectively. Using the $^{29}$Mg mass excess and previous measurements of $^{29}$Cl we uncovered a cubic coefficient of $d$ = 28(7) keV, which is the largest known cubic coefficient of the IMME. This departure, however, could also be caused by experimental data with unknown systematic errors. Hence there is a need to confirm the mass excess of $^{28}$S and the one-neutron separation energy of $^{29}$Cl, which have both come from a single measurement. Finally, our results were compared to ab initio calculations from the valence-space in-medium similarity renormalization group, resulting in a good agreement.
\end{abstract} 

\maketitle

\section{Introduction}

Local atomic mass models have been widely used to predict unknown masses of neutron-deficient nuclei, which are relevant for astrophysical processes such as the rapid-proton capture process \cite{Wre09} and the prediction of long-lived di-proton radioactivity \cite{Bro91}. One of the most robust and successful of such models is the Isobaric Mass Multiplet Equation (IMME) \cite{Wei59, Mac14}. This model makes use of the isospin symmetry in the nucleus that is broken due to the Coulomb interaction. The IMME states that the mass excess (ME) of Isobaric Analogue States (IAS) in a given nuclear multiplet will follow a quadratic dependence with the isospin projection \( T_z = (N - Z)/2 \) of the form:
\begin{equation}
\label{IMME_equation}
\mbox{ME}(A,T,T_z) = a(A,T) + b(A,T)T_z + c(A,T)T_z^2,
\end{equation}
where $a$, $b$, and $c$ are coefficients that depend on the total number of nucleons $A$ in the isobaric group, and the isospin value $T$ of the IAS.

In the past decades, several cases violating the IMME were reported \cite{Her01, Zha12} before more precise measurements of other members of the multiplets re-established the validity of the IMME \cite{Pyl02, Su16}. Among the 35 fully determined quartets and quintets that can test the IMME \cite{Mac14}, only the $A$ = 8 \cite{Cha11}, 9 \cite{Bro12a}, 21 \cite{Gal14}, 31 \cite{Kan16, Ben16}, 32 \cite{Kwi09}, and 35 \cite{Cha07} currently depart enough from quadrature to require the presence of a higher-order term in the IMME. In all these systems, the departure could be justified by the presence of a cubic term of size ranging from -4.3(11) keV \cite{Gal14} to 8.47(14) keV \cite{Mac14}. It is important to know for which isobaric chains additional cubic or quartic terms are required as otherwise the IMME will fail in providing accurate prediction on unknown mass of neutron-deficient nuclei. In this article we report an indirect determination of what would be the largest cubic coefficient in any given multiplet. This determination is based upon new high-precision mass measurements of magnesium isotopes along with recent proton-separation energy measurements of $^{29}$Cl and $^{30}$Ar \cite{Muk15}.

\section{Experimental method}

The atomic masses of $^{27-29}$Mg were measured using TRIUMF's Ion Trap for Atomic and Nuclear science (TITAN) Penning trap mass spectrometer \cite{Dil03, Dil06}, located at the ISAC facility \cite{Dom02, Baa14}. Penning trap mass spectrometry is well established as a tool of choice for precision mass measurements of radioactive isotopes \cite{Bla13}. The radioactive, neutron-rich magnesium isotopes were produced by impinging a 480 MeV, 40 $\mu$A proton beam on a thick SiC target. For our measurements laser resonance ionization in a radiofrequency ion guide (IG-LIS) was chosen. This allows to suppress surface ionized species from the hot target and transfer tube region. Only neutral atoms can effuse into the cold RF ion guide region, where element selective resonant laser ionization occurs \cite{Rae14}.

The TITAN facility currently consists of three ion traps: a radio-frequency quadrupole (RFQ) cooler and buncher \cite{Bru12a}, an electron beam ion trap (EBIT) \cite{Lap12} to charge-breed the ions for increased precision \cite{Ett11,Ett13} as well as to perform in-trap decay spectroscopy \cite{Lea15}, and the mass measurement Penning trap (MPET) \cite{Bro12}. Recently, the system has been used for a series of precision mass measurements with goals ranging from improving nuclear matrix element calculations for double-beta decay experiments \cite{Kwi14} to unveiling discrepancies in reaction rates for the rp-process \cite{Cho15}, and to test new isospin symmetry breaking calculation methods for superallowed pure Fermi transitions \cite{Rei17}. For the neutron-rich magnesium isotope measurements, the mass-selected radioactive beam from the IG-LIS \cite{Rae14} was first accumulated in the RFQ, where it was cooled through collisions with a helium buffer gas and subsequently extracted as bunches. The cooled bunches were then sent directly to MPET as singly charged ions. 

\begin{figure}
\includegraphics[width=8cm]{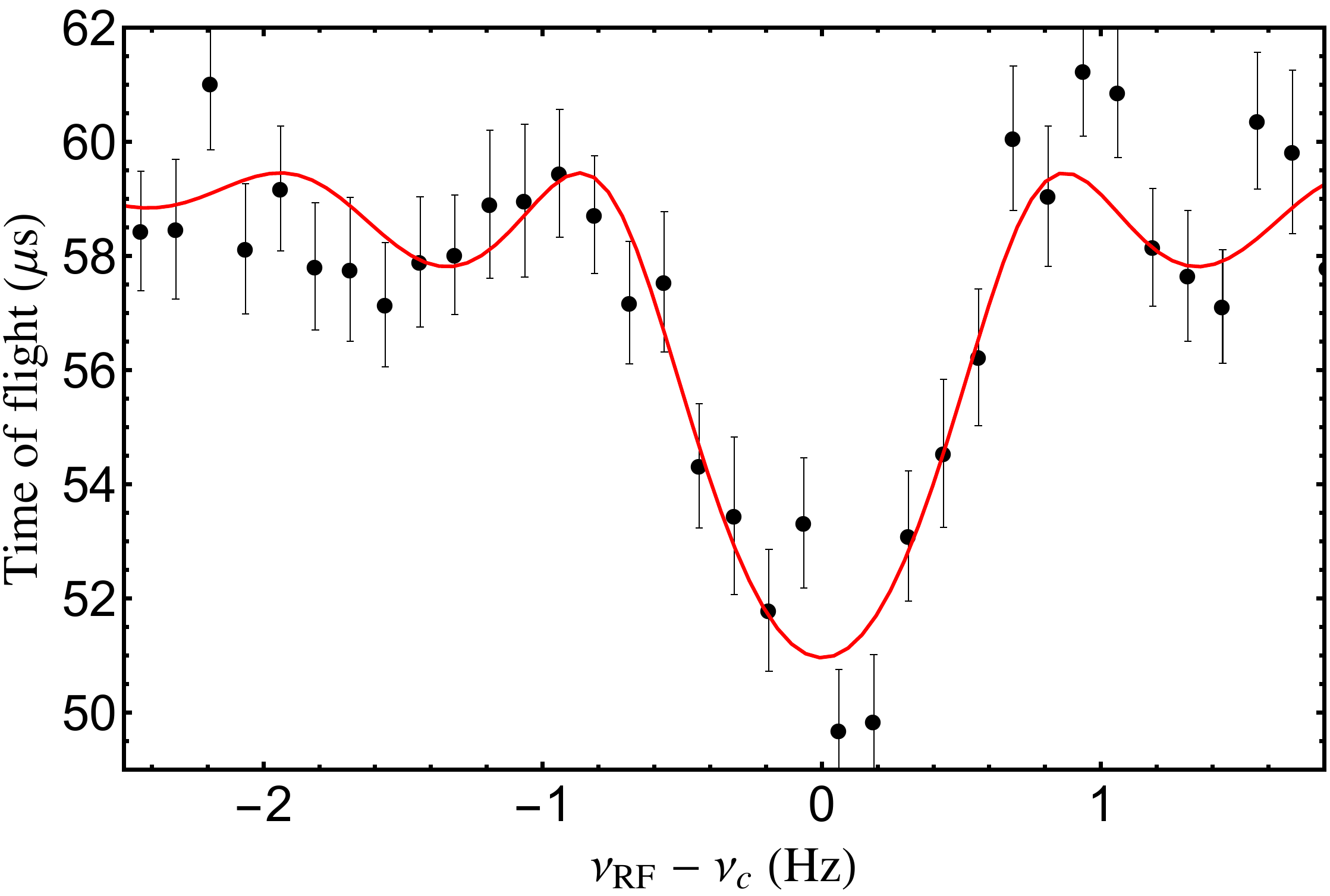}
\caption{(Color on-line) Typical quadrupole resonance of $^{29}$Mg$^+$ with the fitted theoretical line shape \cite{Kon95}. The quadrupole excitation time was $T_{RF}$ = 997 ms.}
\label{29Mgres}
\end{figure}
Once the ions were injected into MPET, the ion cyclotron frequency \(\nu_{c} = qB/(2 \pi m)\) of the ion of interest was determined via the time-of-flight ion cyclotron resonance technique \cite{Gra80, Kon95} (TOF-ICR). A typical quadrupole resonance for the most neutron rich magnesium isotope, $^{29}$Mg, is shown in Figure~\ref{29Mgres}. 

\section{Data analysis}

\begin{table}[t!]
 \caption{\label{table_r} Average of the measured frequency ratios $R$ of $^{27-29}$Mg$^+$ to $^{23}$Na$^+$ utilizing the quadrupole resonance scheme. $N$ indicates the number of measurements taken. The number in parenthesis gives the statistical uncertainty, while the number in brackets gives the total uncertainty.}
 \begin{ruledtabular}
 \begin{tabular}{c c c}
Ref. & ME (keV) & $S_p$ (keV) \\
\hline
\hline
Nuclei & $N$  & Frequency Ratio $\overline{R}$ \\
\hline
$^{27}$Mg & 12 & 1.173 758 460 2(61)[66] \\
\hline
$^{28}$Mg & 3 & 1.217 236 861(12)[12] \\
\hline
$^{29}$Mg & 13 & 1.260 941 332(12)[16] \\
\end{tabular}
\end{ruledtabular}
\end{table}

The atomic masses of our ions of interests were derived from the ratio of the cyclotron frequency of a calibrant ion $\nu_{c,ref}$ over the cyclotron frequency of the ion of interest $\nu_{c}$: $R=\nu_{c,ref}/\nu_{c}$. The cyclotron frequency $\nu_{c,ref}$ is a linear interpolation of two calibrations bracketing the measurement of the ion of interest in order to account for long-term magnetic field fluctuations. 
\begin{figure}
\includegraphics[width=8cm]{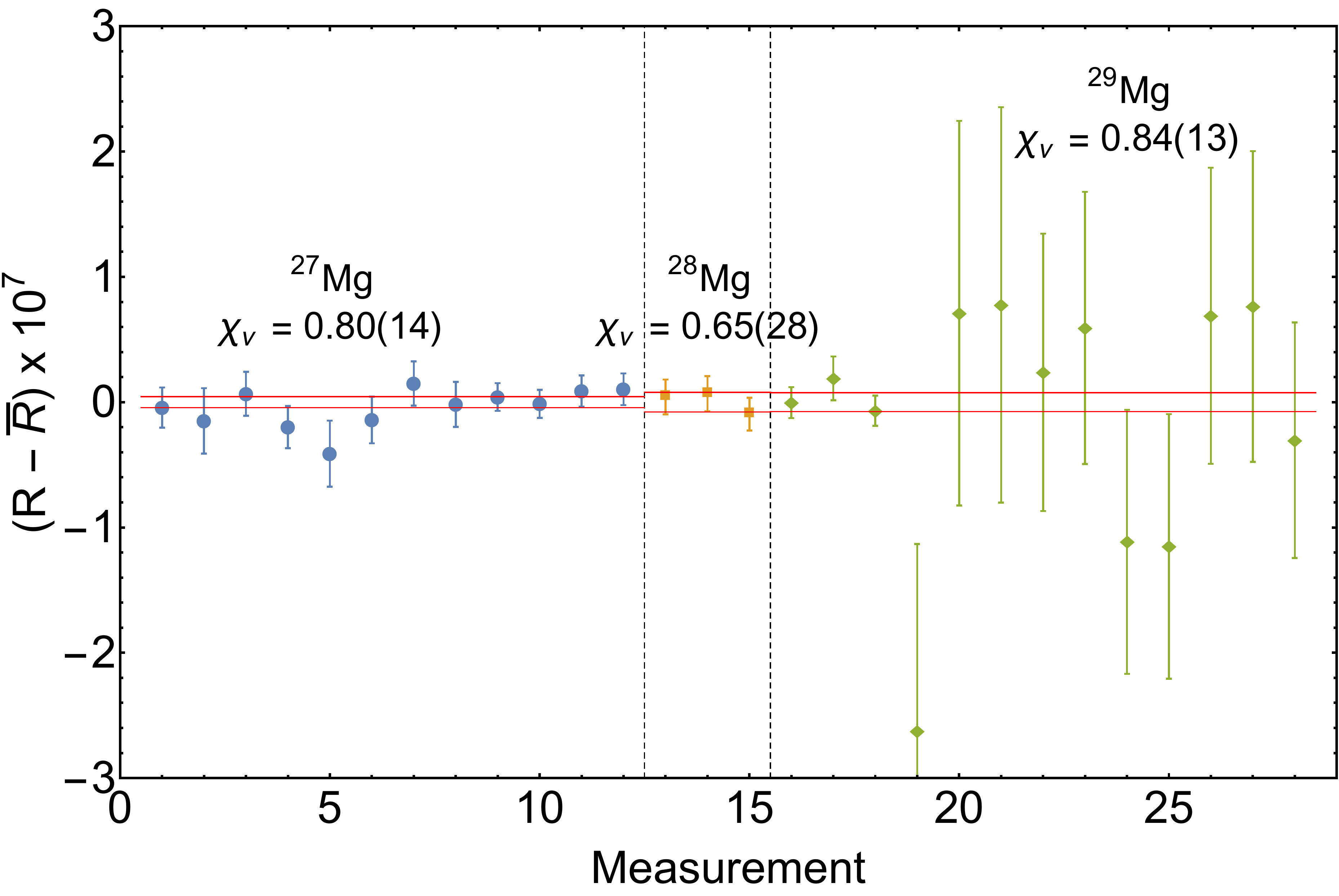}
\caption{(Color on-line) Difference between individual frequency ratios $R$ and the averaged frequency ratios $\overline{R}$ for $^{27-29}$Mg$^+$ ($^{23}$Na$^+$ was the calibrant). The double red lines indicates the 1$\sigma$ statistical uncertainty on $\overline{R}$. The Birge ratio $\chi_{\nu}$ for each weighted average is also given. An excitation time of 997 ms was used for all frequency ratio measurements except the last 10 $^{29}$Mg$^+$ measurements, which were done using a 97 ms excitation.}
\label{FreqRatioPlot}
\end{figure}
Figure~\ref{FreqRatioPlot} shows all the $^{27-29}$Mg$^+$ to $^{23}$Na$^+$ frequency ratios taken. Note that a 997 ms excitation was used for all $^{27-28}$Mg$^+$ measurements and the first three $^{29}$Mg$^+$ measurements. The last ten $^{29}$Mg$^+$ measurements were performed using a 97 ms excitation. The Birge ratios \cite{Bir32} for the three measurements show that the data is statistically distributed. Table~\ref{table_r} gives the weighted average of the frequency ratios.

Most systematic uncertainties in the measured frequency ratios scale linearly with the mass difference between the ion of interest and the calibrant ion \cite{Bro12}. These systematic effects include: magnetic field inhomogeneities, trap misalignment with the magnetic field, harmonic distortion of the electric potential and non-harmonic imperfections in the trapping potential. The combination of these systematic effects was experimentally determined to be $\Delta R/R$ = -0.2(2) ppb $\cdot$ $\Delta A$  for singly charged ions at TITAN \cite{Bro12}. Hence in the data analysis a conservative upper value of 0.4 ppb $\cdot$ $\Delta A$ was added in quadrature as systematic error. The remaining systematic effects stem from non-linear, time-dependent changes in the magnetic field, relativistic effects on the cyclotron frequency, and ion-ion interaction in the trap \cite{Bro09}, which are explained in the following. 

First, the field produced by the superconducting magnet is affected by variations of temperature and pressure. This was minimized at TITAN by using a pressure stabilization system, resulting in temporal magnetic field fluctuations that are typically below 0.15 ppb/h \cite{Bro09}. Because the calibration measurements were separated by less than 75 minutes, the effect of magnetic field variations on the cyclotron frequency are negligible compared to our achieved statistical uncertainty. Second, relativistic effects were also calculated, and found to be negligible compared to our statistical precision, due to the ion's large mass-to-charge ratio \cite{Bro09}. 

Then, the measured cyclotron frequency of the ion of interest can be affected by the presence of contaminant ions of different mass \cite{Bol92}. These ions, which are not affected during the excitation phase, do interact with the ion of interest resulting in a reduction of the measured cyclotron frequency. Possible isobaric contamination was probed by doing a scan of a dipolar excitation \cite{Bla04} at the reduced cyclotron frequency and no contaminant was found. Non-isobaric contaminants  from the off-line ion source or produced by charge-exchange in the RFQ were removed using a Bradbury-Nelson gate \cite{Bru12b} upstream of the Penning trap. Finally, undesirable species can be created from charge-exchange with residual background gas molecules or radioactive decay in the Penning trap itself. Hence, throughout the experiment, we kept the maximum number of recorded ions below two. Nevertheless, an upper value for possible shifts in the frequency ratios due to ion-ion interactions was probed by performing a count-rate analysis \cite{Kel03} on each individual time-of-flight spectrum. In this analysis, cyclotron frequencies extrapolated down to one ion detected were obtained. Because of the small number of counts in each spectrum, a precise estimate of the change in frequency was obtained by taking the difference between the weighted average of frequencies from the count-rate analysis with the uncorrected frequencies. 
\begin{table}[t]
\caption{Weighted average of the cyclotron frequencies obtained from a count-rate analysis $\overline{\nu}_{c,rate}$, together with the weighted average of the uncorrected cyclotron frequencies $\overline{\nu}_c$ and their differences \(\Delta \overline{\nu}_c = \overline{\nu}_{c,rate} - \overline{\nu}_{c}\). The values for the calibrant ($^{23}$Na$^+$), are given below their corresponding Mg isotopes.}
	\begin{tabular}{c  c  c c}
	\hline
	\hline
	Nuclei & $\overline{\nu}_{c,rate}$ (Hz) & $\overline{\nu}_c$ (Hz) & $\Delta \overline{\nu}_c$ (mHz) \\
	\hline 
	$^{27}$Mg$^{+}$ & 2 105 024.578(14) & 2 105  024.584(9) & -6(17) \\
	$^{23}$Na$^{+}$ & 2 470 790.430(12) & 2 470 790.436(7) & -6(14) \\
	\hline
	$^{28}$Mg$^{+}$ & 2 029 835.353(27) & 2 029 835.356(17) & -3(32) \\
	$^{23}$Na$^{+}$ & 2 470 790.437(17) & 2 470 790.443(10) & -6(20) \\
	\hline
	$^{29}$Mg$^{+}$ & 1 959 480.930(39) & 1 959 480.901(17) & 29(43) \\
	$^{23}$Na$^{+}$ & 2 470 790.474(20) & 2 470 790.456(12) & 18(23) \\
	\hline 
	\hline
\end{tabular}
\label{table_ionion}
\end{table}
These frequencies, together with their difference are given in Table \ref{table_ionion}. As it can be seen, all the variation in frequency are within uncertainty. Nevertheless, using the frequency shifts for the Mg isotope $\Delta \overline{\nu}_c$(Mg) and its corresponding calibrant frequency shift $\Delta \overline{\nu}_c$(Na), the relative change in the cyclotron frequency ratio can be estimated using
\begin{equation}
\frac{\Delta R}{R} = \frac{\Delta \overline{\nu}_c(\mbox{Na})}{\overline{\nu}_c(\mbox{Na})} - \frac{\Delta \overline{\nu}_c(\mbox{Mg})}{\overline{\nu}_c(\mbox{Mg})}.
\end{equation}

Finally, previous higher precision measurements using stable isotopes have uncovered a systematic effect amounting to 1.3 ppb due to an unbalanced RF being applied to the Penning trap electrodes \cite{Kwi14}, which has since been resolved.
\begin{table}[t]
\caption{Contribution of the various sources of relative uncertainty on the cyclotron frequency ratios given in part-per-billion (ppb).}
	\begin{tabular}{c c c c}
	\hline
	\hline
	Source & $^{27}$Mg (ppb) & $^{28}$Mg (ppb) & $^{29}$Mg (ppb) \\
	\hline 
	mass-dependent & 1.6 & 2.0 & 2.4 \\
	\hline 
	$B$-field fluctuations & 0.2 & 0.2 & 0.2 \\
	\hline 
	relativistic & 0.2 & 0.2 & 0.2 \\
	\hline 
	ion-ion interaction & 0.4 & 1.0 & 7.5 \\
	\hline 
	unbalanced RF & 1.3 & 1.3 & 1.3 \\
	\hline 
	total systematic & 2.1 & 2.6 & 8.0 \\
	\hline 
	statistical & 5.2 & 9.6 & 9.6 \\
	\hline 
	total uncertainty & 5.6 & 10.0 & 12.5 \\
	\hline 
	\hline
\end{tabular}
\label{table_error}
\end{table}
The contributions of the various forms of uncertainty are summarized in Table \ref{table_error}. As it can be noticed the systematic uncertainty is dominated by our conservative estimate for mass-dependent effects, which is still lower than the statistical uncertainty in the measurements.

\section{Mass excesses}

The atomic mass $m$ of the isotope of interest is determined directly from the averaged cyclotron frequency ratio:
\begin{equation}
m = \overline{R}(m_{ref}-m_{e}) + m_e,
\end{equation}
where $m_{ref}$ is the atomic mass of $^{23}$Na and $m_{e}$ is the electron mass.
\begin{table}[t]
\caption{Mass excess ME of $^{27-29}$Mg from this work (TITAN) and from the latest Atomic Mass Evaluation (\textsc{Ame2016}) \cite{Wan17}. $\Delta$ME is the mass excess difference: $\Delta$ME = ME(TITAN) - ME(\textsc{Ame2016}).}
	\begin{tabular}{c c c c c}
	\hline
	\hline
	Nuclei & ME(TITAN) (keV) & ME(\textsc{Ame2016}) (keV) & $\Delta$ME (keV) \\
	\hline 
	$^{27}$Mg & -14586.38(14) & -14586.611(50) & 0.23(15) \\
	\hline 
	$^{28}$Mg & -15019.95(26) & -15018.8(20) & -1.2(20) \\
	\hline 
	$^{29}$Mg & -10612.38(34) & -10602.8(114) & -9.5(114) \\
	\hline 
	\hline
\end{tabular}
\label{table_ME}
\end{table}
The corresponding mass excess ME, calculated using the $^{23}$Na mass excess from the 2016 Atomic Mass Evaluation (\textsc{Ame2016}) \cite{Wan17}, is presented in Table \ref{table_ME} together with the \textsc{Ame2016} value. The $^{27}$Mg ME is within 1.6$\sigma$ of the \textsc{Ame2016} value, which is based on a series of $^{27}$Mg($n$,$\gamma$) measurements \cite{Wan17}. The $^{28}$Mg ME agrees with the \textsc{Ame2016} value, which is based on a $\beta$-end point measurement \cite{Ols54}, while being 8 times more precise. The $^{29}$Mg ME also agrees with the \textsc{Ame2016} value, which is based on a measurement done with the MISTRAL spectrometer \cite{Gau06}, while being 34 times more precise. 

\section{Discussion}

As discussed in \cite{Bro91}, the mass excess ME($A$,$T$,$-|T_z|$) of a very proton-rich nucleus can be obtained from knowledge of the mass excess, ME($A$,$T$,$T_z$), of its mirror partner using only the $b$ coefficient of the quadratic form of the IMME: 
\begin{equation}
\mbox{ME}(A,T,-|T_z|) =  \mbox{ME}(A,T,T_z) - 2 b(A,T) |T_z|.
\end{equation}
Here we propose, in a new technique, to reverse the situation and use the known mass of a very proton rich nucleus and its mirror partner to probe for a potential cubic term in the quartic form of the IMME:
\begin{equation}
d(A,T) = \frac{\mbox{ME}(A,T,T_z) - 2 b(A,T) |T_z| - \mbox{ME}(A,T,-|T_z|)}{2 |T_z|^3}. 
\end{equation}
This was done using our newly measured $^{29}$Mg mass excess, the mass excess of $^{30}$Mg previously measured at TITAN \cite{Cha13}, which are the neutron-rich mirrors of $^{29}$Cl and $^{30}$Ar, respectively. For this analysis, the mass excess of $^{29}$Cl and $^{30}$Ar were taken from \textsc{Ame2016} \cite{Wan17}. These mass excesses are based on the two-proton separation energy of the drip-line nucleus $^{30}$Ar ($S_{2p}$ = -2.25$^{+0.15}_{-0.10}$ keV), and the one-proton separation energy of its isotonic neighbor $^{29}$Cl ($S_{p}$ = -1.8 $\pm$ 0.1 keV) \cite{Muk15}, which where measured recently. These measurements prompted interest as they indicate that $^{30}$Ar would undergo a very unique type of hybrid decay between a simultaneous emission of two protons and their sequential emission \cite{Muk15}. This unusual type of decay has been argued to be caused by an interplay of three- and two-body decay mechanisms \cite{Muk15}. 

\begin{table*}[t]
\caption{Corrected linear coefficient $b$ estimate from the IMME, using the tabulated $S_b$ and $C_b$ coefficients in the parametrization from \cite{Mac14}, the mass excess ME of $^{30}$Ar, and $^{29}$Cl \cite{Wan17} and corresponding estimated cubic terms.}
	\begin{tabular}{c c c c c c c}
	\hline
	\hline
	Nuclei & ME(\textsc{Ame2016}) (keV) & $S_b$ & $C_b$ (keV) & $b$ (keV) & $d$ (keV) \\
	\hline 
	$^{29}$Cl & 13160(190) & 0.981 & 789 & -4931(22) & 28(7) \\
	\hline 
	$^{30}$Ar & 20930(210) & 1.016 & 1060 & -5057(22) & 10(5) \\
	\hline 
	\hline
\end{tabular}
\label{table_d}
\end{table*}
To extract a reliable cubic term, the $b$ coefficient was calculated using the following parametrization \cite{Mac14}:
\begin{equation}
b_{\mbox{param}}(A,T) = \Delta_{nH} - 720 \cdot S_b(T) \cdot \frac{(A-1)}{A^{1/3}} + C_b(T),
\end{equation}
where $\Delta_{nH}$ = 782.346 64(48) keV is the difference between the mass of the neutron and the proton \cite{Wan17}, 720 keV is obtained from a global fit of the data on IAS, $S_b$(T) is an isospin-dependent correction to that global value, and $C_b$(T) is an isospin-dependent offset \cite{Mac14}. The values of  $S_b$ and $C_b$ for the multiplets of interest are given in Table \ref{table_d}. This $b$ coefficient evaluation was also corrected for the general residuals $\Delta b$ of the parametric fits observed in the region of interest (around $A$ = 30):
\begin{equation}
b(A,T) = b_{\mbox{param}}(A,T) + \Delta b(A).
\end{equation}
\begin{figure}
\includegraphics[width=8cm]{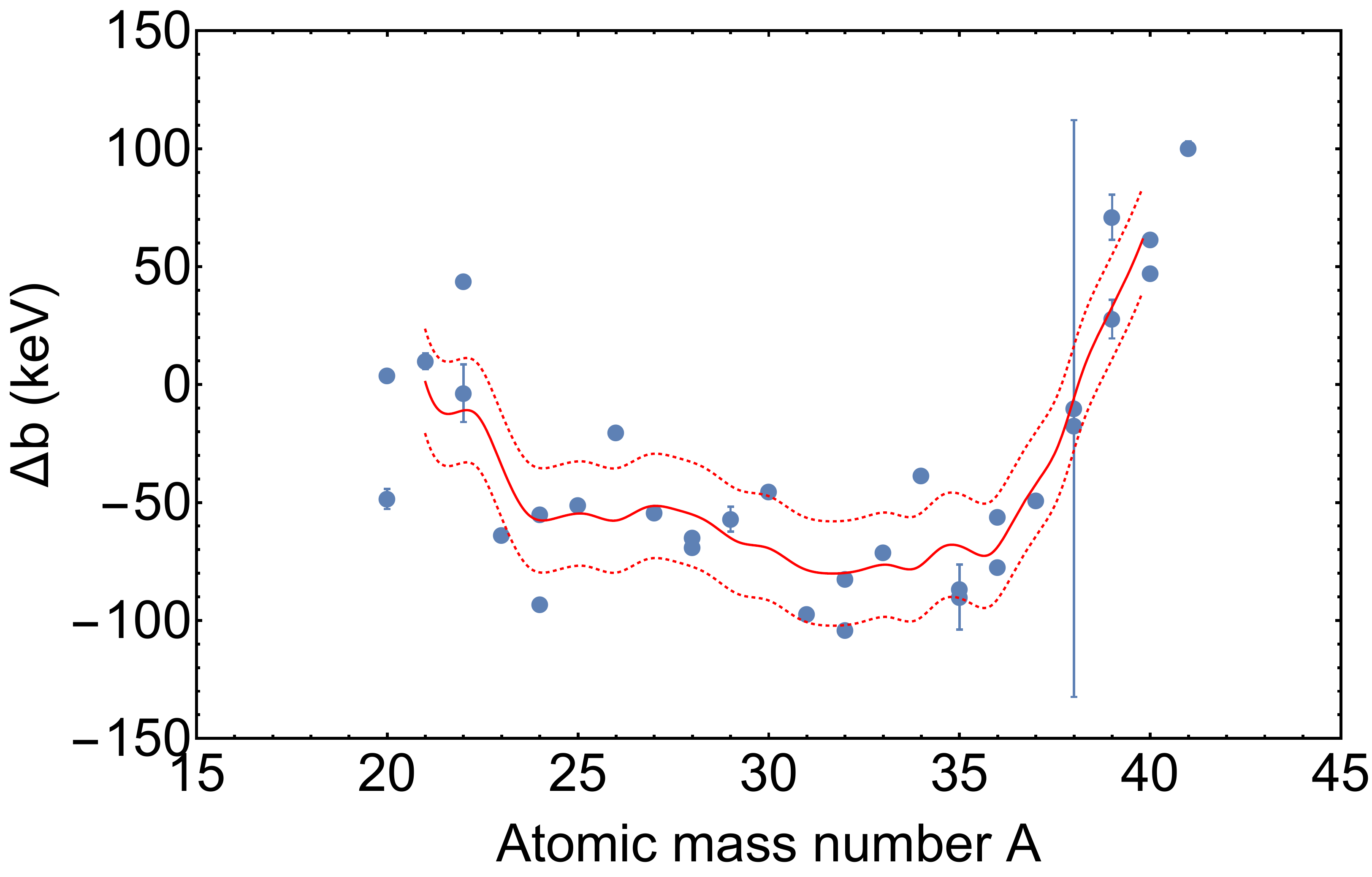}
\caption{(Color on-line) Residuals of the $b$ coefficient parametrization $\Delta b$ as a function of the atomic mass number $A$ for all fully determined multiplets, as described in the text. The solid line represents a spline interpolation of a five-point moving average of the residual. The standard deviation of the difference between the interpolation and the residual is represented by the doted lines.}
\label{bcoeffcalc}
\end{figure}
The residuals of the $b$ coefficient parametrization, together with the spline interpolation of their five-point moving average, are shown in Figure~\ref{bcoeffcalc}. The standard deviation of the difference between the interpolation and the residual was found to be 22 keV and it is indicated by the dotted lines in Figure~\ref{bcoeffcalc}. Note that the spread of the residuals around $A$ = 30 is covered by the 22 keV standard deviation.

\begin{table*}[t]
\caption{Theoretical Mass excess ME of $^{29,30}$Mg, the coefficients $a$, $b$ and $c$ from (\ref{IMME_equation}) and the cubic coefficient $d$  for both interactions and both multiplets, calculated using the valence-space IM-SRG. Additionally the experimental mass excesses ME$_{exp}$ of the nucleus $^{29}$Mg (this work) and $^{30}$Mg (\textsc{Ame2016}) \cite{Wan17} are given, as well as the experimental $b$, $c$ and $d$ coefficients $b_{exp}$, $c_{exp}$ and $d_{exp}$.}
    \begin{tabular}{c c c c c c c c c c c}
    \hline
    \hline
    Interaction & A & ME (keV) & ME$_{exp}$ (keV)  & $a$ (keV) & $b$ (keV) & $b_{exp}$ (keV) & $c$ (keV) & $c_{exp}$ (keV) & $d$ (keV) & $d_{exp}$ (keV)  \\
    \hline 
    SRG 2.0 & 29 & -24901 &  -10612.38(34) & -13586 & -5100 & -4931(22) & 293 & 203(20) & -25 & 28(7) \\
    \hline 
    SRG 1.8 & 29 & -12567 & -10612.38(34) & -1799 & -4978 & -4931(22) & 208 & 203(20) & 24 & 28(7) \\
    \hline 
    SRG 2.0 & 30 & -22921 & -8884(3) & -8280 & -5402 & -5057(22) & 161 & 235(20) & 4 & 10(5) \\
    \hline 
    SRG 1.8 & 30 & -10719 & -8884(3) & 2393 & -4967 & -5057(22) & 187 & 235(20) & 4 & 10(5) \\
    \hline
    \hline
\end{tabular}
\label{table_IMSRG_results}
\end{table*}


The $b$ coefficient, the mass excess of $^{30}$Ar, and $^{29}$Cl, as well as the $d$ coefficient estimate for their respective multiplets are compiled in Table \ref{table_d}. The calculated cubic term for the $A$ = 30, $T$ = 3 multiplet is 10(5) keV, which is 2$\sigma$ from zero and the cubic term for the $A$ = 29, $T$ = 5/2 multiplet is 28(7) keV, which is 4$\sigma$ from zero. The latter cubic term is much greater than any other cubic term observed to date. Such a departure could be a consequence of the Thomas-Ehrman effect observed in unbound systems \cite{Com88}. If this is the case, it would represent the largest shift of this kind, as the cubic terms in lighter systems with unbound nuclei ($A$ = 8, 9) are at most 8.5 keV. 

Another possibility would be that the large cubic terms are due to erroneous experimental data. For example, increasing the mass excess of $^{28}$S, which is used to calculate the mass excess of both $^{30}$Ar and $^{29}$Cl in the \textsc{Ame2016} (since the original data \cite{Muk15} is in the form of neutron separation energies), by 700 keV would re-establish the quadratic behavior of the IMME. Note that the mass excess of $^{28}$S is derived entirely \cite{Wan17} from the $Q$-value measurement of the reaction $^{28}$Si($\pi^+$,$\pi^-$)$^{28}$S \cite{Bur80}, hence a confirmation of the mass excess, preferably using direct means would help solidify the presence of the large cubic coefficient.  

To further shed light on this observation, we calculated the IAS energies of the $A\!=\!29,30$ multiplets from valence-space in-medium similarity renormalization group (IM-SRG) \cite{Tsu12,Bog14,Str16,Str17}, based on two-nucleon (NN) and three-nucleon (3N) forces derived from chiral effective field theory \cite{Epe09,Mac11}. Two different starting chiral Hamiltonians were used in this work. The first (SRG 2.0) uses a chiral NN interaction at next-to-next-to-next-to leading order (N$^3$LO) \cite{Mac03,Mac11} and a 3N force at N$^2$LO \cite{Nav07}, simultaneously evolved with the free-space SRG \cite{Bog07} to a low-momentum scale of $\lambda=2.0\mathrm{fm}^{-1}$ in a single particle spherical harmonic oscillator (HO) basis with energy $\hbar\omega=24$\,MeV.  As discussed in Refs.~\cite{Str16,Str17}, this Hamiltonian reasonably reproduces experimental ground and excited state energies in the lower $sd$ shell.  The second interaction (SRG 1.8), developed in Refs.~\cite{Heb11,Sim16,Sim17}, begins from the same chiral N$^3$LO NN interaction as above, but with $\lambda=1.8\mathrm{fm}^{-1}$.  Unconstrained 3N couplings $c_D$ and $c_E$ are subsequently fit to reproduce the triton binding and alpha particle charge radius at $\Lambda_{\mathrm{3N}}=2.0\mathrm{fm}^{-1}$.  This Hamiltonian, which predicts well saturation energy of infinite symmetric nuclear matter \cite{Heb11}, has recently been shown to reproduce ground-state energies across the nuclear chart from the $p$ shell to the nickel region and beyond \cite{Hag16,Rui16,Sim17,Las17}.  

We then use the Magnus formulation of the IM-SRG \cite{Mor15, Her16} to decouple the $^{16}$O core energy as well as a specific $sd$-shell valence-space Hamiltonian for each nucleus of interest, using the ensemble normal ordering procedure \cite{Str17}, which captures the bulk effects of 3N forces among the valence-space nucleons. These Hamiltonians are diagonalized using the NuShellX shell-model code \cite{NuX} and the IAS are identified using the isospin, parity and angular momentum quantum numbers, obtained from this process. 

Corrections of the isospin mixing in the energies are performed under the assumption of two-level mixing of the states. We calculated the mass excess of the $^{29}$Mg and the $^{30}$Mg ground states, as well as the IAS in $^{29}$Al, $^{29}$S, $^{29}$Cl, $^{30}$Al, $^{30}$Cl and $^{30}$Ar within a HO basis size of $e=2n+l \leq e_{\mathrm{max}} = 12$ and 3N matrix elements where $e_1+e_2+e_3 \leq E_{\mathrm{3max}}=14$. The energies of the IAS were extrapolated in $e_{\mathrm{max}}$ and $E_{\mathrm{3max}}$ and the coefficients of the IMME calculated from the results. This extrapolation procedure was checked for consistency by comparing results to an extrapolation of the coefficients directly calculated from the IAS for increasing $e_{\mathrm{max}}$ and $E_{\mathrm{3max}}$ values. 

Final results are shown in Table \ref{table_IMSRG_results}. The theoretical ME values of the Mg nuclei deviate from experimental data, but results using the SRG 1.8 interaction are in much better agreement.  This is expected since SRG 2.0 is known to overbind nuclei above oxygen isotopes \cite{Bin14}, while SRG 1.8 largely reproduces ground state energies in this region \cite{Sim17}. The IMME coefficients for both interactions are consistent with each other and reproduce well the experimental results. The  $b$ coefficient for SRG 1.8 is even within 100 keV of experiment for both cases, though such an accuracy is well beyond what might be reasonably claimed given unexplored interaction and many-body uncertainties.  For SRG 2.0, $b$ is overestimated on the order of 150 to 350 keV. For completeness, an experimental $c$ coefficient was also calculated using the multiplet-specific fit functions from \cite{Mac14}. The $c$ coefficient calculated from SRG 1.8 for $A$ = 29 is nearly identical to the value based from the fit of all the known multiplets. For $A$ = 30 on the other hand, SRG 1.8 under predict that coefficient by about 50 keV. For both interactions the $d$ coefficient only differs by a few keV from experiment. Considering the fluctuations in the coefficients of lower order and the different signs in the cubic coefficient for $A=29$, we conclude that such agreement is accidental, and the determination of cubic contributions to the IMME is not yet possible at a level of precision comparable to experiment. The consistency of the results, however, highlights the need to further study of the effects of different interactions and to explore estimates of uncertainties from approximations in the many-body approach.  Such work is currently in progress \cite{Dro17}.

\section{Conclusion}

We performed direct mass measurements of neutron-rich magnesium isotopes $^{27-29}$Mg using the TITAN Penning trap mass spectrometer at the ISAC radioactive beam facility at TRIUMF. The $^{27}$Mg mass excess is within 1.6$\sigma$ of the \textsc{Ame2016}, the $^{28,29}$Mg mass excesses are both within 1$\sigma$ of the \textsc{Ame2016}, while being respectively 8 and 34 times more precise. Using the new mass excess of $^{29}$Mg, the previously measured mass excess of $^{30}$Mg at TITAN, and the recent proton-separation energy measurements of $^{29}$Cl and $^{30}$Ar, we calculated the IMME cubic coefficient for the $A$ = 29 and 30 multiplet using a novel approach. For both cases significant non-zero coefficients of 28(7) keV and 10(5) keV, respectively are obtained. In the case of $A$ = 29, the coefficient is much larger than any other cubic term previously obtained experimentally. While ab initio valence-space IM-SRG calculations also find nonzero $d$ coefficients, precision at the tens of keV level is well beyond what can be claimed currently.  Since such a large breakdown of the IMME might be partly caused by inaccurate experimental data, an independent confirmation of the $^{28}$S mass excess, and the $^{29}$Cl one-proton separation energy would help better understand the nature of the large cubic term uncovered.

\section{Acknowledgement}

This work was supported in part by the US National Science Foundation Grant No. PHY-1419765, the Natural Sciences and Engineering Research Council of Canada (NSERC), and the U.S. Department of Energy by Lawrence Livermore National Laboratory under Contract No. DE-AC52-07NA27344. O.M.~Drozdowski gratefully acknowledges financial support from the German Academic Exchange Service (DAAD RISE program). We thank J.~Simonis, K.~Hebeler, and A.~Schwenk for providing the 3N matrix elements used in this work and for valuable discussions. Computations were performed with an allocation of computing resources at the J\"{u}lich Supercomputing Center (JURECA).

\end{document}